\documentclass[aps,prd,twocolumn,floatfix,preprintnumbers,nofootinbib,superscriptaddress]{revtex4}
\usepackage{amsmath,amsthm,amssymb,amsfonts}
\usepackage{graphicx}
\usepackage{textcomp}
\usepackage{bm}
\usepackage{color}
\usepackage{multirow}
\usepackage{dutchcal}
\usepackage{makecell}

\usepackage[outdir=./]{epstopdf}

\begin{document}
\title{Spectrum of the S-wave fully-heavy tetraquark states}

\author{Jie Zhang}
\affiliation{School of Physical Science and Technology, Southwest
University, Chongqing 400715, China}

\author{Jin-Bao Wang}
\affiliation{School of Physics, Southeast University, Nanjing 211189, China}

\author{Gang Li}\email{gli@qfnu.edu.cn}
\affiliation{College of Physics and Engineering, Qufu Normal
University, Qufu 273165, China}

\author{Chun-Sheng An}\email{ancs@swu.edu.cn}
\affiliation{School of Physical Science and Technology, Southwest
University, Chongqing 400715, China}

\author{Cheng-Rong Deng}
\affiliation{School of Physical Science and Technology, Southwest
University, Chongqing 400715, China}

\author{Ju-Jun Xie}
\affiliation{Institute of Modern Physics, Chinese Academy of
Sciences, Lanzhou 730000, China} \affiliation{School of Nuclear
Science and Technology, University of Chinese Academy of Sciences,
Beijing 101408, China} \affiliation{Lanzhou Center for Theoretical Physics, Key
Laboratory of Theoretical Physics of Gansu Province, Lanzhou
University, Lanzhou 730000, China}

\date{\today}

\begin{abstract}
	
In present work, spectrum of the $S$-wave fully-heavy tetraquark states $QQ\bar{Q}\bar{Q}$ ($Q=c,b$), i.e., $cc\bar{c}\bar{c}$, $bb\bar{b}\bar{b}$, $cc\bar{b}\bar{b}$/$bb\bar{c}\bar{c}$, $bc\bar{c}\bar{c}$/ $cc\bar{b}\bar{c}$, $bb\bar{c}\bar{b}$/$cb\bar{b}\bar{b}$, and $bc\bar{b}\bar{c}$ are systematically investigated within a nonrelativistic constituent quark model, in which the instanton-induced and one-gluon-exchange interactions are taken into account as the residual spin-dependent hyperfine interaction. Our results show that the states with $cc\bar{c}\bar{c}$ and $bb\bar{b}\bar{b}$ components could be located around $ 6500$~MeV and $ 19200$~MeV, respectively. Based on our calculations, the new $X(6900)$ state observed by LHCb may be not a ground $cc\bar{c}\bar{c}$ tetraquark state, while it could be an orbitally or radially excited state of $cc\bar{c}\bar{c}$ system. On the other hand, the recently reported $X(6600)$ state by CMS and ATLAS can be explained as a ground $cc\bar{c}\bar{c}$ tetraquark state with spin-parity $J^{PC} =0^{++}$.
	
\end{abstract}

\maketitle

\section{Introduction}
\label{intro}

Structure and properties of the $XYZ$ states, those are beyond the traditional $q\bar{q}$ constituent quark picture, have been intensively studied in the last two decades~\cite{Chen:2016qju,Lebed:2016hpi,Guo:2017jvc,Liu:2019zoy,Brambilla:2019esw,Chen:2022asf}, since the discovery of the famous $X(3872)$~\cite{Belle:2003nnu} in the exclusive~$B^{\pm}\rightarrow K^{\pm}\pi^{+}\pi^{-}J/\psi$ decays. Theoretically, the observed $XYZ$ states are often described as compact tetraquark states, hadronic molecular states, or admixtures of these two kinds of states. Alternatively, one can also treat some of the exotic candidates to be admixtures of the $q\bar{q}$ and $(q\bar{q})^2$ components. In any case, systematical investigations on the meson exotic states can of course deepen our understanding on the non-perturbative regime of QCD theory.

Among the meson exotic states, tremendous investigations on the fully-heavy tetraquark states have been taken very recently, inspired by the experimental measurement carried by LHCb that a broad structure in the range $6.2-6.8$~GeV and a narrow structure at $\sim6.9$ GeV are observed in the di-$J/\psi$ invariant spectrum~\cite{LHCb:2020bwg}, which indicates the existence of the all-charm tetraquark states. In fact, theoretical investigations on the fully-heavy tetraquark states could be traced back to 1980s~\cite{Chao:1980dv,Ader:1981db,Heller:1985cb,Badalian:1985es,Zouzou:1986qh}, while these works have not gotten enough attentions because of lack of experimental data. In the new century, discovery of more and more meson exotic states attracted attentions of hadronic physicists again on the fully-heavy tetraquark states, lots of theoretical predictions on spectrum of the $QQ\bar{Q}\bar{Q}$~($Q=c,b$) tetraquark states have been done before 2020~\cite{Lloyd:2003yc,Barnea:2006sd,Berezhnoy:2011xn,Chen:2016jxd,Wu:2016vtq,Karliner:2016zzc,Bai:2016int,Wang:2017jtz,Richard:2017vry,Debastiani:2017msn,Anwar:2017toa,Hughes:2017xie,Esposito:2018cwh,Wang:2019rdo,Liu:2019zuc}.  Since the discovery of $X(6900)$ in 2020~\cite{LHCb:2020bwg}, tremendous theoretical investigations on the structure, production and decay of all-heavy tetraquark states have been taken using different approaches such as the constituent quark model, diquark-diquark picture, QCD sum rules, effective field theory, etc.~\cite{Jin:2020jfc,Lu:2020cns,Chen:2020xwe,Faustov:2020qfm,Zhao:2020nwy,Weng:2020jao,Zhu:2020xni,Zhao:2020zjh,Yang:2021zrc,Li:2021ygk,Yang:2021hrb,Tiwari:2021tmz,Wang:2021mma,Liu:2021rtn,Esposito:2021ptx,Zhuang:2021pci,Asadi:2021ids,Kuang:2022vdy,Wu:2022qwd,Hu:2022zdh,Yang:2022cut,Gong:2022hgd,Wang:2022xja,Chen:2022sbf,An:2022qpt,Mutuk:2022nkw,Li:2019uch,Li:2018bkh,Zhou:2022xpd}. Both the compact and hadronic molecular structure have been considered in these theoretical investigations, and most of the results suggest that the experimentally observed $X(6900)$ may be not a ground state of the tetraquark/molecular state. In addition to the LHCb experiments~\cite{LHCb:2020bwg}, the CMS~\cite{CMS6600} and ATLAS~\cite{atlas6600} collaborations have reported their measurements on the fully-charm structure at the very recent ICHEP 2022 conference. And three all-charm structures are reported, whose Breit-Wigner masses are~\cite{CMS6600}:
\begin{eqnarray}
M_X(6600) &=& 6552 \pm 10 \pm 12 ~{\rm MeV}\,,\nonumber\\
M_X(6900) &=& 6927 \pm 9 \pm 5 ~{\rm MeV}\,,\nonumber\\
M_X(7300) &=& 7287 \pm 19 \pm 5 ~{\rm MeV}\,. \nonumber
\end{eqnarray}

Besides the all-charm tetraquark states, experimental efforts on searching the beauty-full tetraquark states $bb\bar{b}\bar{b}$ have also been made by years. The CMS collaboration have suggested a possible energy region for searching the $bb\bar{b}\bar{b}$ structure in the proton-proton collision  recently~\cite{CMS:2016liw}, but there is no confirmed signature found by later measurements carried by LHCb and CMS collaborations~\cite{LHCb:2018uwm,CMS:2020qwa}.

Accordingly, it's still worthy for us to study the fully-heavy tetraquark states systematically. As we know, the constituent quark model is one of the most powerful approach to investigate on spectrum of the multi-quark states. In a very recent work~\cite{Wang:2022clw}, the constituent quark model with an instanton-induced effective interaction has been developed to study the spectrum of $S$-wave hidden and double charm tetraquark states, and some interesting results consistent with the experimental data and other theoretical investigations have been obtained. Along this line, in present work, we study the mass spectra of the $S-$wave fully-heavy tetraquark states $QQ\bar{Q}\bar{Q}$ using the quark model as shown in Ref.~\cite{Wang:2022clw}. In addition, here we also employ the widely used one-gluon-exchange model to estimate energies of the fully-heavy tetraquark states, and discuss the numerical results obtained with the two different approaches.

The present manuscript is organized as follows. In Sec.~\ref{Theory}, we show the theoretical formalism of this work, including effective Hamiltonian and explicit wave functions of the studied fully-heavy tetraquark states. Determinations of all the model parameters and the corresponding numerical results for spectrum of fully-heavy tetraquark states in the presently employed instanton-induced and one-gluon-exchange models are presented in Sec.~\ref{ResultsAndDiscussions}. Finally, Sec.~\ref{Summary} is a brief summary of the present work.

\section{Framework} \label{Theory}

\subsection{Effective Hamiltonian}
\label{hal}

Here we employ a nonrelativistic quark potential model to study the mass spectra of the fully-heavy tetraquark states. Accordingly, the effective Hamiltonian can be expressed as follow
\begin{equation}
H_{eff}=\sum_{i=1}^{4}\left(m_i+T_i\right)-T_{C.M.}+V_{conf}+V_{hyp}^{cont}\,,
\label{H}
\end{equation}
where $m_i$ and $T_i$ denote the constituent mass and kinetic energy of $i$-th quark, respectively, the second term $-T_{C.M.}$ is to remove contributions from motions of the center of mass. $V_{Conf}$ stands for the quark confinement potential, in present model, we adopt the form~\cite{Wang:2022clw,Cornell}:
\begin{equation}
V_{conf}=\sum_{i<j}-\frac{3}{16}\,\left(\vec{\lambda}^c_i\cdot\vec{\lambda}^c_j\right)\,\left(b\,r_{ij}-\frac{4}{3}\frac{\alpha_{ij}}{r_{ij}}+C_0\right)\,,
\label{confine}
\end{equation}
where $\vec{\lambda}^c$ is the $SU(3)$ Gell-mann matrix in color space, and the first term in the second bracket provides the long-range linear interaction, the second term provides the short-range Coulomb interaction, and the last term is the zero point energy. $b$ and $\alpha_{ij}$ are the long-range coupling strength of quark confinement, QCD effective coupling constant between a quark pair, respectively.

The last term $V_{hyp}^{cont}$ in Eq.~(\ref{H}) is the residual (contact part) spin-dependent hyperfine interaction that should cause the mass splittings of different flavor-spin-color configurations, for which the most widely accepted versions are the one-gluon-exchange model (OGE), the Goldstone-bose-exchange model (GBE) and the instanton-induced interaction (INS). The OGE model has been intensively applied to study mass spectra of both the traditional hadrons and the multiquark states. While the INS model has also been employed to reproduce the spectrum of the light hadrons and the single charm baryons~\cite{Brau:1998sxe,Semay:2001th,Migura:2006ep}, as well the tetraquark states with light quarks~\cite{Beinker:1995qe,tHooft:2008rus}. And the INS model has also been considered to investigate on the $S$-wave hidden- and double-charmed tetraquark states very recently~\cite{Wang:2022clw}. On the other hand, the GBE interaction between a quark-antiquark pair in a multiquark system is often assumed to vanish~\cite{Helminen:2000jb,Duan:2016rkr,An:2017lwg,Wang:2021rjk}.
Here, for the hyperfine potential $V^{cont}_{hyp}$, we take into account the OGE or the INS interactions separately.

In the OGE model, the spin-dependent contact hyperfine interaction which cause the mass splitting is taken as the following form~\cite{Capstick:2000qj}:
\begin{equation}
V_{OGE}^{cont} = -\frac{\pi}{6} \sum_{i<j}\alpha_{ij} \left(\vec{\lambda}^c_i\cdot\vec{\lambda}^c_j\right) \frac{\vec\sigma_i\cdot\vec\sigma_j}{m_im_j}\,\delta^3\left(\vec{r}_{ij}\right)\,,
\end{equation}
where $\sigma_i$ and $\sigma_j$ are the Pauli matrices. While the hyperfine interaction in the INS model is categorized into two parts~\cite{Beinker:1995qe,Wang:2022clw}:
\begin{equation}
V_{INS}^{cont}=V^{cont}_{INS}(qq)+V^{cont}_{INS}(q\bar{q}) \,,  \label{vins}
\end{equation}
with $V^{cont}_{INS}(qq)$ and $V^{cont}_{INS}(q\bar{q})$ standing for the interactions between a quark-quark pair and a quark-antiquark pair, respectively. The explicit forms for them are
\begin{widetext}
\begin{eqnarray}
V^{cont}_{INS}(qq) &=&  -\sum_{i<j}\hat{g}^{qq}_{ij}\left(P_{ij}^{S=1}P_{ij}^{C,\bf{6}}+2P_{ij}^{S=0}P_{ij}^{C,\bf{\bar{3}}}\right)\delta^3\left(\vec{r}_{ij}\right)\,,\\
V^{cont}_{INS}(q\bar{q})  &=&  \sum_{i<j}\hat{g}^{q\bar{q}}_{ij}\left[\frac{3}{2}P_{ij}^{S=1}P_{ij}^{C,\bf{8}}+P_{ij}^{S=0}\left(\frac{1}{2}P_{ij}^{C,\bf{8}} + 8P_{ij}^{C,\bf{1}}\right)\right]\delta^3\left(\vec{r}_{ij}\right)\,,
\end{eqnarray}
\end{widetext}
where $\hat{g}^{qq}_{ij}$ and $\hat{g}^{q\bar{q}}_{ij}$ are flavor-dependent coupling strength operators, explicit matrix elements of these operators in the light, strange and charm quark sectors have been given in Ref.~\cite{Wang:2022clw}, and the results including bottom quark are similar.
$P_{ij}^{S=0}$ and $P_{ij}^{S=1}$ are spin projector operators onto spin-singlet and spin-triplet states, respectively,
and $P_{ij}^{C,\bf{\bar{3}}}$, $P_{ij}^{C,\bf{6}}$ and $P_{ij}^{C,\bf{8}}$ are color projector operators onto color anti-triplet $\bf{\bar{3}}_c$, color sextet $\bf{6}_c$ and color octet $\bf{8}_c$, respectively.
Note that one should keep in mind that the interaction between two antiquarks is of the same form as $V^{qq}_{INS}$, if one changes all colours to anti-colours and all flavours to their anti-flavour indices.

Note that the presently employed hyperfine interaction is pure contact interaction, which should lead to an unbound Hamiltonian if we don't treat it perturbatively. Thus we need to regularize it as in Refs.~\cite{Beinker:1995qe,Wang:2022clw},
\begin{equation}
\delta^3\left(\vec{r}_{ij}\right)\,\rightarrow\,\left(\frac{\sigma}{\sqrt{\pi}}\right)^3\mathrm{exp}\left(-\sigma^2\,r_{ij}^2\right)\,.
\end{equation}
The regularization parameter $\sigma$ should be correlated to the finite size of constituent quarks and therefore flavor dependent, but here we just make $\sigma$ to be a unified parameter for all the flavors, and absorb the flavor dependence into the coupling strengths, to reduce the free model parameters.

\subsection{Wave functions}
\label{wfcs}

Presently, we consider the $S$-wave fully-heavy $QQ\bar{Q}\bar{Q}$ tetraquark states, including the $cc\bar{c}\bar{c}$, $bb\bar{b}\bar{b}$, $cc\bar{b}\bar{b}$/$bb\bar{c}\bar{c}$, $bc\bar{c}\bar{c}$/$cc\bar{b}\bar{c}$, $bc\bar{b}\bar{b}$/$bb\bar{b}\bar{c}$, and $bc\bar{b}\bar{c}$ states. Hereafter, we will not list the conjugate states such as $bb\bar{c}\bar{c}$. In general, the flavor configurations for the $QQ\bar{Q}\bar{Q}$ system can be:
\begin{flalign}
\hspace{1cm}|1\rangle&=\{QQ\}_{{\bf{6}}_c}\{\bar{Q}\bar{Q}\}_{\bar{\bf{6}}_c}\,, & |2\rangle&=[QQ]_{\bar{\bf{3}}_c}[\bar{Q}\bar{Q}]_{{\bf{3}}_c}\,,\hspace{1cm}\notag\\
|3\rangle&=\{QQ\}^{*}_{\bar{\bf{3}}_c}[\bar{Q}\bar{Q}]_{{\bf{3}}_c}\,, & |4\rangle&=[QQ]_{\bar{\bf{3}}_c}\{\bar{Q}\bar{Q}\}^{*}_{{\bf{3}}_c}\,,\notag\\
|5\rangle&=[QQ]^{*}_{{\bf{6}}_c}\{\bar{Q}\bar{Q}\}_{\bar{\bf{6}}_c}\,, & |6\rangle&=\{QQ\}_{{\bf{6}}_c}[\bar{Q}\bar{Q}]^{*}_{\bar{\bf{6}}_c}\,,\notag\\
|7\rangle&=\{QQ\}^{*}_{\bar{\bf{3}}_c}\{\bar{Q}\bar{Q}\}^{*}_{{\bf{3}}_c}\,, & |8\rangle&=[QQ]^{*}_{{\bf{6}}_c}[\bar{Q}\bar{Q}]^{*}_{\bar{\bf{6}}_c}\,,
\end{flalign}
where $\{\cdots\}$ denotes the permutation symmetric flavor state while $[\cdots]$ the anti-symmetric one for a two-quark (-antiquark) system. With or without the superscript $*$ shows the spin-triplet or spin-singlet. The blackened number in the subscript denotes corresponding color symmetry of two-quark subsystem.
Accordingly, we present all the possible flavor configurations with appropriate quantum numbers $J^{PC}$ in Table~\ref{HCSym}.

\begin{table}[ht]
\caption{Flavor configurations of fully-heavy tetraquark states.}
\label{HCSym}
\renewcommand
\tabcolsep{0.55cm}	
\renewcommand{\arraystretch}{1.4}
\begin{tabular}{ccc}
\hline\hline
		
Flavor content &   $J^{PC}$ &  Configurations\\\hline
		
\multirow{3}{*}{$cc\bar{c}\bar{c}$}&$0^{++}$&$\begin{matrix}|1\rangle\\|7\rangle\end{matrix}$\\
\cline{2-3}
		
		&$1^{+-}$&$|7\rangle$\\
		
		&$2^{++}$&$|7\rangle$\\\hline

\multirow{3}{*}{$bb\bar{b}\bar{b}$}&$0^{++}$&$\begin{matrix}|1\rangle\\|7\rangle\end{matrix}$\\
\cline{2-3}

		&$1^{+-}$&$|7\rangle$\\
		&$2^{++}$&$|7\rangle$\\\hline
\multirow{3}{*}{$cc\bar{b}\bar{b}$}&$0^{++}$&$\begin{matrix}|1\rangle\\|7\rangle\end{matrix}$\\
\cline{2-3}

		&$1^{+-}$&$|7\rangle$\\
		&$2^{++}$&$|7\rangle$\\\hline
\multirow{6}{*}{$cc\bar{b}\bar{c}$}&$0^{+}$&$\begin{matrix}|1\rangle\\|7\rangle\end{matrix}$\\
\cline{2-3}

		&$1^{+}$&$\begin{matrix}|3\rangle\\|6\rangle\\|7\rangle\end{matrix}$\\
		\cline{2-3}

		&$2^{+}$&$|7\rangle$\\\hline
		
\multirow{6}{*}{$bb\bar{b}\bar{c}$}&$0^{+}$&$\begin{matrix}|1\rangle\\|7\rangle\end{matrix}$\\
\cline{2-3}

        &$1^{+}$&$\begin{matrix}|3\rangle\\|6\rangle\\|7\rangle\end{matrix}$\\
\cline{2-3}
        &$2^{+}$&$|7\rangle$\\\hline		
\multirow{10}{*}{$bc\bar{b}\bar{c}$}&$0^{++}$&$\begin{matrix}|1\rangle\\|2\rangle\\|7\rangle\\|8\rangle\end{matrix}$\\
\cline{2-3}

		&\multirow{2}{*}{$1^{++}$}&$|3'\rangle=\frac{1}{\sqrt{2}}\left(|3\rangle+|4\rangle\right)$\\

		&        &$|5'\rangle=\frac{1}{\sqrt{2}}\left(|5\rangle+|6\rangle\right)$\\	
\cline{2-3}		
		&\multirow{4}{*}{$1^{+-}$}&$|4'\rangle=\frac{1}{\sqrt{2}}\left(|3\rangle-|4\rangle\right)$\\	
		&        &$|6'\rangle=\frac{1}{\sqrt{2}}\left(|5\rangle-|6\rangle\right)$\\	
		&        &$|7\rangle$\\	
		&        &$|8\rangle$\\
\cline{2-3}		
		&$2^{++}$&$\begin{matrix}|7\rangle\\|8\rangle\end{matrix}$\\ 	

\hline\hline	
\end{tabular}
\end{table}

To get the explicit matrix elements of the Hamiltonian, one has to solve the Shr\"{o}dinger equation of the Hamiltonian in Eq.~(\ref{H}). Here we employ the Gaussian functions expansion method (GFEM)~\cite{GEM}, which has been applied to study the spectrum of petaquark states in Ref.~\cite{Zhang:2005jz} and $QQ\bar{q}\bar{q}$ tetraquark states in Ref.~\cite{Zhang:2007mu}, and spectrum of the $S$-wave hidden- and double-charm tetraquark states in Ref.~\cite{Wang:2022clw} very recently.

In GFEM, one can use a series of Gaussian functions to expand the orbital wave function of the $S$-wave tetraquark system as follow
\begin{equation}
\Psi(\{\vec{r}_i\})=\prod_{i=1}^{4}\sum_{l}^{n}C_{il}\left(\frac{1}{\pi b_{il}^2}\right)^{3/4}\mathrm{exp}\left[-\frac{1}{2b_{il}^2}r_i^2\right]\,,
\end{equation}
where the parameter $\{b_{il}\}$ can be related to the harmonic oscillator frequencies $\{\omega_{l}\}$ by $1/b_{il}^2=m_i\omega_{l}$. In order to facilitate the calculations, we take the ansatz that $\omega_{l}$ is independent to the quark mass as in Ref.~\cite{Liu:2019zuc}, i.e., $1/b_{il}^2=1/b_{l}^2=m_l\omega_{l}$, then the spatial wave function can be simplified to be
\begin{align}
\Psi(\{\vec{r}_i\})=&\sum_{l}^{n}C_{l}\prod_{i=1}^{4}\left(\frac{m_i\omega_{l}}{\pi }\right)^{3/4}\mathrm{exp}\left[-\frac{m_i\omega_{l}}{2}r_i^2\right]\notag\\
=&\sum_{l}^{n}C_{l}\,\psi\left(\omega_{l},\{\vec{r}_i\}\right)\, .\label{SpatialF}
\end{align}
In addition, to remove contributions of the motion of the center of mass directly, here we used the Jacobi coordinates $\{\vec{r}_{i}\}$ those are defiend as
\begin{eqnarray}
  \vec{\xi}_{1} &=& \vec{r}_{1}-\vec{r}_{2}\,,\nonumber \\
  \vec{\xi}_{2} &=& \vec{r}_{3}-\vec{r}_{4}\,, \nonumber\\
  \vec{\xi}_{3} &=& \frac{m_1 \vec{r}_{1}+m_2\vec{r}_{2}}{m_1+m_2}- \frac{m_3 \vec{r}_{3}+m_4\vec{r}_{4}}{m_3+m_4}\,,\nonumber\\
  \vec{R} &=&  \frac{m_1 \vec{r}_{1}+m_2\vec{r}_{2}+m_3 \vec{r}_{3}+m_4\vec{r}_{4}}{m_1+m_2+m_3+m_4}\,,
\end{eqnarray}
where we have enumerated the two quarks by $1,2$ and the two antiquarks by $3,4$, respectively.

Following the routine in Ref.~\cite{GEM}, the parameters $\{b_{l}\}$ are set to be geometric series,
\begin{equation}
b_{l}=b_1a^{l-1}\hspace{0.8cm}\left(l=1,2,...,n\right)\,,
\end{equation}
where $n$ is the number of Gaussian functions and $a$ the ratio coefficient, then there are only thee parameters $\{b_{1},b_{n},n\}$ need to be determined.

Now we are able to compute the diagonal enlements of the Hamiltonian matrixes through solving the generalized matrix eigenvalue problem, using the patial wave function Eq.~(\ref{SpatialF}), the flavor$\,\times\,$spin$\,\times\,$color wave functions introduced in Eq., and Hamiltonian Eq.~(\ref{H}),
\begin{equation}
\sum_{l}^n\sum_{l'}^nC^i_{l}\left(H^d_{ll'}-E_i^d N_{ll'}\right)C^i_{l'}=0\,,\label{GEP}
\end{equation}
where $i=1\text{--}n$ and
\begin{align}
H^d_{ll'}=&\langle\psi\left(\omega_{l}\right)(FSC)|H_{eff.}|\psi\left(\omega_{l'}\right)(FSC)\rangle\,,\\
N_{ll'}=&\langle\psi\left(\omega_{l}\right)(FSC)|\psi\left(\omega_{l'}\right)(FSC)\rangle\,,
\end{align}
here $(FSC)$ stand for flavor$\,\otimes\,$spin$\,\otimes\,$color wave function. One can choose a set of $\{b_{1},b_{n},n\}$ tentatively, then extend and densify the harmonic oscillator length parameters until the minimum energy $E^d_m$, which should be correlative to the energy of psysical state according to the Rayleigh-Ritz variational principle, becomes stable. In present work we take $\{b_{1},b_{n},n\}$=$\{0.02$fm$,6$fm$,40\}$. At last, the nondiagonal elements of Hamiltonian matrices can be easily obtained using $\{C_{l}^m\}$.

\section{Numerical results and discussions}  \label{ResultsAndDiscussions}

\subsection{Model parameters}  \label{paras}

To get the numerical results, a reliable set of parameters is required. The parameters in present model are the constituent quark masses, quark confinement strength $b$, QCD effective coupling constants $\{\alpha_{ff^{\prime}}\}$ and the coupling strengths in INS hyperfine interaction $\{g_{ff^{\prime}}\}$ for a two quark pair, the zero point energy $C_0$, and the regulation parameter of the $\delta$ function $\sigma$. Here the constituent quark masses $m_n$, $m_s$ and $m_c$ and corresponding coupling strengths are just taken to be the same values used in~\cite{Wang:2022clw}, for the constituent quark mass of the beauty quark $m_b$, the empirical value is in the range $4900-5100$~MeV~\cite{Liu:2019zuc,Yang:2022cut}, here we take it to be $5000$~MeV tentatively.

Since there is no solid experimental data for the spectrum of pure exotic states, thus it's impossible for us to fix the parameters by fitting the masses of tetraquark states. On the other hand, the presently employed quark confinement and hyperfine interactions are just the two-body interactions, namely, same as the description of traditional baryons and mesons in the constituent quark model. Consequently, here we try to give the values for present model parameters by fitting the masses of a set of the traditional ground state hadrons listed in PDG~\cite{ParticleDataGroup:2020ssz}, using the INS and OGE models, respectively.

\begin{table}[htbp]
\caption{Model parameters.}
\label{ModelParameters}
\renewcommand
\tabcolsep{0.52cm}
\renewcommand{\arraystretch}{1.8}
\begin{tabular}{ccc}
\hline\hline
& INS  &  OGE \\
\hline
$m_n\,\,(\mathrm{MeV})$   &  $340$  &   $340$   \\
$m_s\,\,(\mathrm{MeV})$   &  $511$  &   $511$   \\
$m_c\,\,(\mathrm{MeV})$   &  $1674$  &   $1674$   \\
$m_b\,\,(\mathrm{MeV})$   &  $5000$  &   $5000$   \\
$b\,\,(\mathrm{10^3MeV^2})$   &  $77$    &   $115$ \\
$C_0\,\,(\mathrm{MeV})$   &  $-296$  &   $-301$  \\

$\alpha_{nn}$             &  $0.60$  &  $0.63$ \\
$\alpha_{ns}$             &  $0.60$  &  $0.62$ \\
$\alpha_{ss}$             &  $0.56$  &  $0.51$ \\
$\alpha_{nc}$             &  $0.51$  &  $0.53$ \\
$\alpha_{sc}$             &  $0.48$  &  $0.49$ \\
$\alpha_{cc}$             &  $0.45$  &  $0.55$ \\
$\alpha_{bb}$             &  $0.40$  &   $0.44$  \\
$\alpha_{bc}$             &  $0.43$  &   $0.48$  \\

$g_{nn}\,\,(\mathrm{10^{-6}MeV^{-2}})$ & $16.9$ &  $-$  \\
$g_{ns}\,\,(\mathrm{10^{-6}MeV^{-2}})$ & $10.7$ &   $-$  \\
$g_{nc}\,\,(\mathrm{10^{-6}MeV^{-2}})$ & $4.0$ &   $-$  \\
$g_{sc}\,\,(\mathrm{10^{-6}MeV^{-2}})$ & $3.1$ &   $-$  \\
$g_{bc}\,\,(\mathrm{10^{-6}MeV^{-2}})$ & $0.58$ &  $-$  \\
$\sigma\,\,(\mathrm{MeV})$      &  $485$   & $1500$ \\
\hline\hline

\end{tabular}
\end{table}


\begin{table*}[htbp]
\caption{ Ground state hadron spectrum in the presently employed INS and OGE models compared to the experimental data, in unit of MeV.}
\label{FittingResults}
\renewcommand
\tabcolsep{0.20cm}
\renewcommand{\arraystretch}{1.8}
\begin{tabular}{cccccccccccccccccc}
\hline\hline
Meson & $\pi$ & $\rho$ & $\omega$ & $\phi$ & $K$ & $K^*$ & $D$ & $D^{*}$ & $D_{s}$ & $D_s^*$
& $\eta_c$ & $J/\psi$ & $B^+_c$ & $B^{*+}_c$ & $\eta_b$ & $\Upsilon$ \\

$PDG$ & $138$ & $775$ & $783$  &  $1019$ &  $496$  & $895$  &  $1867$  &  $2009$  &  $1968$ & $2112$ &  $2983$  &  $3097$ & $6274$   & $-$ &  $9399$  &  $9460$ \\

$INS$ & $139$ & $775$ & $775$  &  $1019$ &  $499$  & $894$  &  $1866$  &  $2009$  &  $1969$ & $2111$ &  $-$   &  $3097$ & $6274$   & $6328$ & $-$    &  $9461$ \\

$OGE$ & $138$ & $785$ & $785$  &  $1019$ &  $496$  & $888$  &  $1867$  &  $1957$  &  $2064$ & $2112$ &  $2984$  &  $3097$ & $6279$   & $6329$ &  $9396$  &  $9430$ \\

\hline
Baryon & $\Sigma_c$ & $\Sigma^*_c$ & $\Xi_c$  &  $\Xi_c^\prime$ &  $\Xi_c^*$  & $\Omega_c$ & $\Omega_c^*$  &   $\Xi_{cc}$\\

$PDG$ & $2454$ & $2518$ & $2469$  &  $2577$ &  $2646$  & $2695$ & $2766$  &  $3621$\\

$INS$ & $2447$ & $2518$ & $2507$  &  $2554$ &  $2627$  & $2664$ & $2737$  &  $3607$\\

$OGE$ & $2449$ & $2488$ & $2534$  &  $2580$ &  $2611$  & $2722$ & $2746$  &  $3601$\\

\hline\hline
\end{tabular}
\end{table*}

All the model parameters (used or not for estimation on the spectrum of fully-heavy tetraquark states in present work) in each model are collected in Table~\ref{ModelParameters}, and the corresponding fitting results compared to the experimental data are shown in Table~\ref{FittingResults}. As we can see in the tables, the experimental data can be well reproduced in both the INS and OGE models with a set of reasonable values for the model parameters. One may note that the $\eta_{c}$ and $\eta_{b}$ mesons In INS model have not been considered in the fitting procedure. This is because of that the INS interaction vanishes between a quark-antiquark with the same flavor, and this should yield unreasonable numerical results that the $\eta_{c(b)}$ and $J/\psi(\Upsilon)$ are degenerated states. To solve this problem, one has to consider mixing between the pseudoscalar mesons $q\bar{q}$ with the quark and antiquark being the same flavor naturally caused by the instanton interaction~\cite{tHooft:1999cta}, while there is not solid experimental data for this kind of mixing angles, therefore, these meson states have not been taken into account. While masses of all the other hadrons obtained in INS model are perfectly consistent with the experimental data, as we can see in Table~\ref{FittingResults}.

Straightforwardly, one can calculate the energies of ground $S-$wave fully-heavy tetraquark systems with the model parameters in Table~\ref{ModelParameters}. The final numerical results for the $cc\bar{c}\bar{c}$ and $bb\bar{b}\bar{b}$ states, $cc\bar{b}\bar{b}$ states and $bc\bar{b}\bar{c}$ states, and $cc\bar{b}\bar{c}$ and $bb\bar{b}\bar{c}$ states are presented in Table~\ref{4c4b}, Table~\ref{2c2b} and Table~\ref{3plus1}, respectively. In the following three subsections, we will discuss the numerical results of the different kinds of fully-heavy tetraquark states respectively.


\subsection{The $cc\bar{c}\bar{c}$ and $bb\bar{b}\bar{b}$ states} \label{dis4c4b}

The numerical results for the spectrum of $cc\bar{c}\bar{c}$ and $bb\bar{b}\bar{b}$ states withe quantum numbers $J^{PC}=0^{++}$, $1^{+-}$ and $2^{++}$ in both the presently employed INS and OGE models are presented in Table~\ref{4c4b}, compared to some of the predictions by quark models~\cite{Ader:1981db,Lloyd:2003yc,Wu:2016vtq,Karliner:2016zzc,Liu:2019zuc}. As we can see in the table, the $S$-wave $cc\bar{c}\bar{c}$ and $bb\bar{b}\bar{b}$ states falls in the range $\sim6400-6500$~MeV and $\sim19200$~MeV, respectively. This is in general consistent with most of the theoretical predictions by various of theoretical approaches. For instance, in the early work~\cite{Ader:1981db}, a quark model with a two-body potential due to color-octet exchange were employed to study the all-charm tetraquark states, and the obtained energies of the $S$-wave $cc\bar{c}\bar{c}$ states were $\sim6400$~MeV. In a recent work base on the QCD sum rule framework~\cite{Chen:2016jxd}, the predicted $cc\bar{c}\bar{c}$ states fall in the range $6400-6800$~MeV. And the quark models predict the $S$-wave $bb\bar{b}\bar{b}$ states fall in the range $18700-18900$~MeV in Refs.~\cite{Karliner:2016zzc,Anwar:2017toa}, but $19300$~MeV in Ref.~\cite{Liu:2019zuc}.

\begin{table*}[htbp]
\caption{The numerical results of spectrum for the $S$-wave $cc\bar{c}\bar{c}$ and $bb\bar{b}\bar{b}$ tetraquark states, where the columns $M_{INS}$ and $M_{OGE}$ show the numerical results of the obtained energies in unit of MeV empoying the INS and OGE models, respectively, one should note that the eigenvalues are obtained by taking into account effects of the configurations mixing, and the columns $C_{INS}$ and $C_{OGE}$ present the mixing coefficients of the corresponding flavor-spin-color configurations in INS and OGE models, respectively. Finally, in the last five columns, we compare the presently obtained results with the predictions in Refs.~\cite{Ader:1981db,Lloyd:2003yc,Wu:2016vtq,Karliner:2016zzc,Liu:2019zuc} using the quark models.}
\label{4c4b}
\renewcommand
\tabcolsep{0.20cm}
\renewcommand{\arraystretch}{1.8}
\begin{tabular}{c|ccccccccccc}
\hline\hline

Systems&$J^{PC}$  &  Config.  &   $M_{INS}$   & $C_{INS}$  &   $M_{OGE}$   &  $C_{OGE}$ & Ref.\cite{Ader:1981db} & Ref.~\cite{Lloyd:2003yc}   & Ref.~\cite{Wu:2016vtq} &Ref.~\cite{Karliner:2016zzc} & Ref.~\cite{Liu:2019zuc}\\
\hline

\multirow{4}{*}{$cc\bar{c}\bar{c}$} & \multirow{2}{*}{$0^{++}$} & $|1\rangle$ & $6414$ & $(1,\hspace{0.15cm}0)$ & $6500$ & $(-0.81,\hspace{0.15cm}-0.58)$ & $6383$ & $6695$ & $7016$ & $6192$& $6518$ \\

                                    &  & $|7\rangle$  & $6414$ & $(0,\hspace{0.15cm}1)$
                                    &  $6411$  &  $(0.58,\hspace{0.15cm}-0.81)$ & $6437$ & $6477$ & $6797$ &$-$ & $6487$\\
\cline{2-12}
                                   &$1^{+-}$  &  $|7\rangle$  &   $6414$  & $(1)$ &  $6453$ & $(1)$ & $6437$ & $6528$& $6899$&$-$& $6500$\\
\cline{2-12}
                                &$2^{++}$   &   $|7\rangle$  &   $6414$  &  $(1)$ & $6475$ &$(1)$ &$6437$& $6573$ & $6956$& $-$& $6524$ \\
\hline
\multirow{4}{*}{$bb\bar{b}\bar{b}$}& \multirow{2}{*}{$0^{++}$}  & $|1\rangle$  & $19226$ & $(1,\hspace{0.15cm}0)$   &  $19235$  &  $(-0.81,\hspace{0.15cm}0.58)$ & $-$ &$-$ & $20275$& $18826$& $19338$\\

&  & $|7\rangle$  & $19226$ & $(0,\hspace{0.15cm}1)$
                                    &  $19200$  &  $(-0.58,\hspace{0.15cm}-0.81)$ &$-$ &  $-$ & $20155$& $-$&$19332$\\
\cline{2-12}
                             &$1^{+-}$&$|7\rangle$  &  $19226$ &  $(1)$ &  $19216$  & $(1)$ &$-$ & $-$ & $20212$&$-$& $19329$\\
\cline{2-12}
                              &$2^{++}$&$|7\rangle$  &  $19226$  &  $(1)$ & $19225$  & $(1)$ &$-$ & $-$ & $20243$&$-$& $19341$\\

\hline\hline
\end{tabular}
\end{table*}

Especially, in a recent work~\cite{Liu:2019zuc}, the fully-heavy $QQ\bar{Q}\bar{Q}$ states has been investigated using the constituent quark model with the OGE interaction, which is almost the same as the presently employed OGE model with only a few differences in the details of the fitting and calculation procedure. Numerical results for spectrum of the $S-$wave $cc\bar{c}\bar{c}$ and $bb\bar{b}\bar{b}$ states in~\cite{Liu:2019zuc} are in general close to but a little bit higher than the presently obtained ones in the OGE model. It's because of that a zero point energy $C_0=-301$~MeV in the quark confinement potential is considered in present OGE model, although the constituent masses of the charm and beauty quark are taken to be higher values than in Ref.~\cite{Liu:2019zuc}.

For the $cc\bar{c}\bar{c}$ states, the obtained energies in both the INS and OGE models suggest that the experimentally observed $X(6900)$ by LHCb~\cite{LHCb:2020bwg} may not be the $S-$wave tetraquark state, but a radially or orbitally excited state, since the obtained energies are $\sim400$~MeV lower than $X(6900)$. And the CMS reported $X(7300)$~\cite{CMS6600} is $\sim800$~MeV higher than the presently obtained states. While the presently obtained energies just fall in the range of the broad structure reported by LHCb~\cite{LHCb:2020bwg}.

In addition, the obtained energy $\sim6500$~MeV of the $cc\bar{c}\bar{c}$ states is very close to the very recently reported $X(6600)$ state by CMS  and ATLAS collaborations~\cite{CMS6600,atlas6600}, therefore, one may expect that the $X(6600)$ state should be dominated by the $S-$wave $cc\bar{c}\bar{c}$ tetraquark component. While in a very recent work~\cite{Wang:2022yes}, the lowest $cc\bar{c}\bar{c}$ resonant state is predicted to be located at $\sim7050$~MeV. And the all-charm tetraquark states are studied in the framework of QCD sum rule in~\cite{Wang:2022xja}, where the $X(6600)$ state is assigned to be the first radially excited $cc\bar{c}\bar{c}$ state with $J^{PC}=0^{++}$. In Ref.~\cite{Wang:2022jmb}, it's shown that contributions of four characteristic intermediate channels $\eta_{c}\chi_{c1}$, $J/\psi\psi(3686)$, $\chi_{c0}\chi_{c1}$ and $\chi_{c2}\chi_{c2}$ are required to reproduce the CMS distribution.

Concerning the other components in $X(6600)$, the di-$J/\psi$ molecular bound state with quantum number $J^{PC}=0^{++}$ or $2^{++}$ whose threshold is just $\sim200-300$~MeV lower than the presently obtained $cc\bar{c}\bar{c}$ states should be expected. Generally, mixing between the tetraquark state and the molecular state should raise up energy of the higher state which is dominated by the $cc\bar{c}\bar{c}$. Consequently, an admixture of the presently obtained $S-$wave $cc\bar{c}\bar{c}$ component~(dominant one) and a di-$J/\psi$ molecular component may account for the experimentally observed $X(6600)$. And one may also expect that the presently obtained $cc\bar{c}\bar{c}$ state with $J^{PC}=1^{+-}$ should couple to the $\eta_{c}J/\psi$ molecular state. Therefore, it's difficult to determine the quantum number of $X(6600)$ by present model, although $J^{PC}=0^{++}$ may be the most prefered one, since energies of the states with different quantum numbers are close.

Similarly, one may also expect the mixing between the $bb\bar{b}\bar{b}$ states with $J^{PC}=0^{++}$ or $2^{++}$ in present model and the di-$\Upsilon$ molecular state, and the states with $J^{PC}=1^{-+}$ and $\eta_{b}\Upsilon$ molecular state, since energies of the presently obtained states are very close to either the di-$\Upsilon$ threshold $18920$~MeV, or the $\eta_{b}\Upsilon$ threshold $18859$~MeV. While there is no convincing experimental signature for candidates of the exotic beauty-full state up to now, hope the future LHCb and CMS experiments could provide more information on this kind of exotic meson states.

On the other hand, as one can see in Table~\ref{4c4b}, the obtained $cc\bar{c}\bar{c}$ and $bb\bar{b}\bar{b}$ states with different quantum numbers are degenerated in the INS model, additionally, there is no mixing between different flavor-spin-color configurations. But the mixing effects in the OGE model are very strong. It's because of that the INS interaction vanishes in these configurations, thus only the quark confinement potential which doesn't cause the mass splittings contributes to the energies. While mass splittings about several tens~MeV for the states with different quantum numbers are obtained in the OGE model.
Tentatively, to distinguish the states with different quantum numbers in INS model, one can roughly estimate the mixing of $cc\bar{c}\bar{c}$ states with different quantum numbers and the recently predicted hidden charm tetraquark states $c\bar{c}n\bar{n}$ and $c\bar{c}s\bar{s}$ in Ref.~\cite{Wang:2022clw} naturally caused by the INS interactions between the quark-antiquark pairs.
The different mixing coefficients should result in the non-degenerate state, but the mixing effects cannot be very strong~\cite{Wang:2022clw}. On the other hand, one can also take the mixing between the presently studied $b\bar{b}b\bar{b}$ states and the $b\bar{b}n\bar{n}$ and $b\bar{b}s\bar{s}$ states in the INS model, which should cause the mass splittings of the $b\bar{b}b\bar{b}$ states with different quantum numbers.

\begin{table*}[htbp]
\caption{The numerical results of spectrum for the $S-$wave $cc\bar{b}\bar{b}$ and $bc\bar{b}\bar{c}$ states compared to the predictions in Ref.~\cite{Liu:2019zuc}, conventions are the same as in Table~\ref{4c4b}.}
\label{2c2b}
\renewcommand
\tabcolsep{0.25cm}
\renewcommand{\arraystretch}{1.8}
\begin{tabular}{c|ccccccc}
\hline\hline

Systems&$J^{PC}$  &  Config.  &   $M_{INS}$   & $C_{INS}$  &   $M_{OGE}$   &  $C_{OGE}$  & CQM~\cite{Liu:2019zuc}\\
\hline

\multirow{4}{*}{$cc\bar{b}\bar{b}$} & \multirow{2}{*}{$0^{+}$} & $|1\rangle$ & $12893$ & $(-0.92,\hspace{0.15cm}-0.40)$ & $12981$ & $(-0.98,\hspace{0.15cm}0.21)$ & $13039$ \\
                                       &  & $|7\rangle$  & $12791$ & $(0.40,\hspace{0.15cm}-0.92)$ &  $12880$  &  $(-0.21,\hspace{0.15cm}-0.98)$ & $12947$\\
\cline{2-8}
                                       &$1^{+}$ & $|7\rangle$  &  $12818$ &  $(1)$ &  $12890$  & $(1)$ & $12960$\\
\cline{2-8}
                                       &$2^{+}$ & $|7\rangle$ & $12838$    &  $(1)$ & $12902$ &$(1)$ &   $12972$ \\
\hline
\multirow{12}{*}{$bc\bar{b}\bar{c}$} & \multirow{4}{*}{$0^{++}$} & $|1\rangle$ & $12887$ & $(-0.61,\hspace{0.15cm}0.48,\hspace{0.15cm}-0.55,\hspace{0.15cm}-0.30)$ & $12966$ & $(0.52,\hspace{0.15cm}-0.54,\hspace{0.15cm}-0.56,\hspace{0.15cm}0.36)$ & $13050$  \\
                 &  & $|2\rangle$ & $12837$ & $(0.35,\hspace{0.15cm}-0.43,\hspace{0.15cm}-0.83,\hspace{0.15cm}0.09)$ & $13035$ & $(0.38,\hspace{0.15cm}-0.54,\hspace{0.15cm}0.67,\hspace{0.15cm}-0.31)$ & $12864$ \\
                 &  & $|7\rangle$ & $12790$ & $(0.67,\hspace{0.15cm}0.73,\hspace{0.15cm}-0.09,\hspace{0.15cm}-0.03)$ & $12850$ & $(-0.68,\hspace{0.15cm}-0.54,\hspace{0.15cm}0.18,\hspace{0.15cm}0.46)$ & $12864$ \\
                 &  & $|8\rangle$ & $12774$ & $(-0.21,\hspace{0.15cm}0.22,\hspace{0.15cm}-0.09,\hspace{0.15cm}0.95)$ & $12783$ & $(0.34,\hspace{0.15cm}0.36,\hspace{0.15cm}0.44,\hspace{0.15cm}0.75)$  & $12835$\\
\cline{2-8}
& \multirow{2}{*}{$1^{++}$} & $|3'\rangle$ & $12850$ & $(-0.86,\hspace{0.15cm}-0.50)$ & $12938$ & $(-0.82,\hspace{0.15cm}-0.58)$  & $13056$\\
                         &  & $|5'\rangle$ & $12794$ & $(0.50,\hspace{0.15cm}-0.86)$ &  $12851$  &  $(0.58,\hspace{0.15cm}-0.82)$ & $12870$ \\
\cline{2-8}
&\multirow{4}{*}{$1^{+-}$} &$|4'\rangle$ & $12886$ & $(0.75,\hspace{0.15cm}0.66,\hspace{0.15cm}0,\hspace{0.15cm}0)$ & $12964$ & $(0.77,\hspace{0.15cm}0.62,\hspace{0.15cm}0.12,\hspace{0.15cm}-0.06)$  & $13052$\\
 &  & $|6'\rangle$ & $12865$ & $(0,\hspace{0.15cm}0,\hspace{0.15cm}0.94,\hspace{0.15cm}-0.34)$ & $12949$ & $(0.11,\hspace{0.15cm}0.08,\hspace{0.15cm}-0.85,\hspace{0.15cm}0.52)$  & $13047$ \\
 &  & $|7\rangle$ & $12794$ & $(0.66,\hspace{0.15cm}-0.75,\hspace{0.15cm}0,\hspace{0.15cm}0)$ & $12835$ & $(-0.45,\hspace{0.15cm}0.56,\hspace{0.15cm}0.36,\hspace{0.15cm}0.60)$  & $12864$ \\
 &  & $|8\rangle$ & $12789$ & $(0,\hspace{0.15cm}0,\hspace{0.15cm}0.34,\hspace{0.15cm}0.94)$ & $12802$ & $(0.43,\hspace{0.15cm}-0.55,\hspace{0.15cm}0.38,\hspace{0.15cm}0.61)$  & $12852$ \\
\cline{2-8}
& \multirow{2}{*}{$2^{++}$} & $|7\rangle$ & $12896$ & $(-0.84,\hspace{0.15cm}0.54)$ & $12964$ & $(-0.82,\hspace{0.15cm}0.58)$  & $13070$ \\
&  & $|8\rangle$ & $12795$ & $(-0.54,\hspace{0.15cm}-0.84)$ &  $12852$  &  $(-0.58,\hspace{0.15cm}-0.82)$ & $12864$ \\

\hline\hline
\end{tabular}
\end{table*}

\subsection{The $cc\bar{b}\bar{b}$, $bc\bar{b}\bar{c}$, $cc\bar{b}\bar{c}$ and $bb\bar{b}\bar{c}$ states}
\label{others}

As we can see in Tables~\ref{2c2b} and ~\ref{3plus1}, the $cc\bar{b}\bar{b}$ and $bc\bar{b}\bar{c}$ states fall in the range $\sim12850\pm50$~MeV, and the $cc\bar{b}\bar{c}$ and $bb\bar{b}\bar{c}$ states are located at $\sim9650\pm50$~MeV and $\sim16050\pm50$~MeV, respectively. All the numerical results are in general consistent with the predictions by other theoretical investigations.

Similar to the case of the $cc\bar{c}\bar{c}$ and $bb\bar{b}\bar{b}$ states, the obtained energies in the INS model are in general lower than those obtained in the OGE model. Note that since the zero point energy $C_0$ is considered in the present work, the numerical results in both the INS and OGE models are about several tens (or up to 100)~MeV lower than the predicted values in Ref.~\cite{Liu:2019zuc} where the OGE interactions are considered in the constituent quark model.

While in the INS model, it's different from the $cc\bar{c}\bar{c}$ and $bb\bar{b}\bar{b}$ states, the obtained  $cc\bar{b}\bar{b}$, $bc\bar{b}\bar{c}$, $cc\bar{b}\bar{c}$ and $bb\bar{b}\bar{c}$ states with different quantum numbers are non-degenerate since the INS interactions survive in these states, in addition, mixing between the different flavor-spin-color configurations is also caused by INS interactions between quarks. And in both of the INS and OGE models, the configuration mixing effects are very strong, as shown in columns $C_{INS}$ and $C_{OGE}$ of the Tables~\ref{2c2b} and~\ref{3plus1}.

Note that the $bc\bar{b}\bar{c}$ states should also mix with the $cc\bar{c}\bar{c}$ or $bb\bar{b}\bar{b}$ states in the INS model. While the mixing should be expected to be very small because of that the coupling strength $g_{bc}$ is about one order smaller than $g_{nc}$ and $g_{ns}$. Thus here we don't discuss much about this kind of mixing. While mixing between the presently tetraquark states and the meson-meson molecular states, for instance, the mixing of the $bc\bar{b}\bar{c}$ states with proper quantum numbers and $J/\psi\Upsilon$ or $\eta_{c}\Upsilon$ should be more interesting in future theoretical investigations.

\begin{table*}[htbp]
\caption{The numerical results of spectrum for the $S$-wave $cc\bar{b}\bar{c}$ and $bb\bar{b}\bar{c}$ states compared to the predictions in Ref.~\cite{Liu:2019zuc}, conventions same as in Table~\ref{4c4b}.}
\label{3plus1}
\renewcommand
\tabcolsep{0.40cm}
\renewcommand{\arraystretch}{1.8}
\begin{tabular}{c|ccccccc}
\hline\hline
Systems&$J^{PC}$  &  Config.  &   $M_{INS}$   & $C_{INS}$  &   $M_{OGE}$   &  $C_{OGE}$ & CQM~\cite{Liu:2019zuc}\\
\hline

\multirow{6}{*}{$cc\bar{b}\bar{c}$} & \multirow{2}{*}{$0^{+}$} & $|1\rangle$ & $9652$ & $(-0.78,\hspace{0.15cm}0.63)$ & $9732$ & $(-0.81,\hspace{0.15cm}-0.58)$ & $9763$ \\
                                &  & $|7\rangle$  & $9615$ & $(-0.63,\hspace{0.15cm}-0.78)$ &  $9665$  &  $(0.58,\hspace{0.15cm}-0.81)$ & $9740$ \\
\cline{2-8}
& \multirow{3}{*}{$1^{+}$} & $|3\rangle$ & $9646$ & $(-0.05,\hspace{0.15cm}-0.63,\hspace{0.15cm}0.77)$ & $9718$ & $(0.56,\hspace{0.15cm}-0.80,\hspace{0.15cm}0.21)$ & $9757$ \\
&  &  $|6\rangle$ & $9630$ & $(0.76,\hspace{0.15cm}0.48,\hspace{0.15cm}0.43)$ & $9699$ & $(-0.44,\hspace{0.15cm}-0.07,\hspace{0.15cm}0.90)$ & $9749$ \\
&  &  $|7\rangle$ & $9605$ & $(0.65,\hspace{0.15cm}-0.61,\hspace{0.15cm}-0.46)$ & $9676$ & $(-0.71,\hspace{0.15cm}-0.59,\hspace{0.15cm}-0.39)$ & $9746$ \\
\cline{2-8}
 &$2^{+}$ & $|7\rangle$  &   $9645$  &  $(1)$ & $9713$ &$(1)$ & $9768$  \\
\hline
\multirow{6}{*}{$bb\bar{b}\bar{c}$} & \multirow{2}{*}{$0^{+}$} & $|1\rangle$ & $16059$ & $(-0.78,\hspace{0.15cm}0.63)$ & $16100$ & $(-0.81,\hspace{0.15cm}0.58)$  & $16173$\\
&  & $|7\rangle$  & $16019$ & $(-0.63,\hspace{0.15cm}-0.78)$ &  $16061$  &  $(-0.58,\hspace{0.15cm}-0.81)$ & $16158$\\
\cline{2-8}
& \multirow{3}{*}{$1^{+}$} & $|3\rangle$ & $16053$ & $(-0.05,\hspace{0.15cm}0.63,\hspace{0.15cm}0.78)$ & $16089$ & $(0.53,\hspace{0.15cm}0.83,\hspace{0.15cm}0.18)$ & $16167$ \\
&  &  $|6\rangle$ & $16036$ & $(0.76,\hspace{0.15cm}-0.48,\hspace{0.15cm}0.44)$ & $16079$ & $(-0.26,\hspace{0.15cm}-0.05,\hspace{0.15cm}0.96)$ & $16164$ \\
&  &  $|7\rangle$ & $16009$ & $(-0.65,\hspace{0.15cm}-0.61,\hspace{0.15cm}0.46)$ & $16046$ & $(0.81,\hspace{0.15cm}-0.56,\hspace{0.15cm}0.19)$ & $16157$ \\
\cline{2-8}
 &$2^{+}$ & $|7\rangle$  &   $16051$  &  $(1)$ & $16089$ &$(1)$ & $16176$ \\
\hline\hline
\end{tabular}
\end{table*}

\section{Summary}
\label{Summary}

In present work, the spectrum of the $S$-wave fully-heavy tetraquark states $cc\bar{c}\bar{c}$, $bb\bar{b}\bar{b}$, $bb\bar{c}\bar{c}$/$cc\bar{b}\bar{b}$, $bc\bar{c}\bar{c}$/$cc\bar{b}\bar{c}$, $bb\bar{c}\bar{b}$/$cb\bar{b}\bar{b}$, and $bc\bar{b}\bar{c}$ are investigated systematically in a nonrelativistic constituent quark model, within which the instanton-induced and one-gluon-exchange interactions are taken into account respectively as the residual spin-dependent hyperfine interaction between quark pairs those cause the mass splittings for different flavor-spin-color configurations. The model parameters in both the two employed models are determined by fitting them to the mass spectrum of the traditional $q\bar{q}$ and $qqq$ hadron states.

The theoretical calculations here show that the ground $cc\bar{c}\bar{c}$ state is with mass around $6500$~MeV, thus the experimentally observed $X(6900)$ state may not be an $S$-wave $cc\bar{c}\bar{c}$ tetraquark state, but it could be a radially or orbitally excited state of $cc\bar{c}\bar{c}$. In addition, the very recently reported all-charm structure $X(7300)$ also cannot be explained by the $S$-wave tetraquark state. This is in agreement with most of the previous predictions by various of other approaches.

While the presently studied $cc\bar{c}\bar{c}$ tetraquark state with quantum number $J^{PC}=0^{++}$, color configuration of the quark-quark and antiquark-antiquark being $\bf{6_c}\otimes\bar{\bf{6}}_c$, and both the quark pair and antiquark pair being the spin singlet, may be the dominant component in the very recently reported $X(6600)$ state by CMS and ATLAS collaborations.

\begin{acknowledgments}

This work is partly supported by the Chongqing Natural Science Foundation under Project Nos. cstc2021jcyj-msxmX0078 and cstc2019jcyj-msxmX0409, and the National Natural Science Foundation of China under Grant Nos. 12075288, 12075133, 11735003, 11961141012 and 11835015. It is also
supported by the Youth Innovation Promotion Association CAS, Taishan Scholar Project of Shandong Province (Grant No.tsqn202103062), the Higher Educational Youth Innovation Science and Technology Program Shandong Province (Grant No. 2020KJJ004).

\end{acknowledgments}

\end{document}